\documentclass{nature}
\usepackage{textcomp}
\usepackage[a4paper,top=2cm,bottom=2cm,left=2.cm,right=2.cm]{geometry}
\usepackage{epsfig}
\usepackage{lineno}

\usepackage{emptypage}
\usepackage{babel}
\usepackage[bf, footnotesize]{caption}
\usepackage{color}
\usepackage{amsmath}
\usepackage{amssymb}
\usepackage[T1]{fontenc}
\usepackage[dvipsnames]{xcolor}
\definecolor{gray}{gray}{0.25}
\usepackage{lastpage}
\usepackage{fancyhdr}
\usepackage{hyperref}
\usepackage{lineno}
\usepackage{caption}
\usepackage{subcaption}

\newcommand{\hi}{H\,\textsc{i}}

\newcommand{\apx}{$\sim$}

\newcommand{\target}{B2\,0258+35}

\newcommand{\kmps}{km~s$^{-1}$}
\newcommand{\p}[1]{$^{-#1}$}
\newcommand{\pp}[1]{$^{#1}$}

\newcommand{\halpha}{H$\alpha$}

\newcommand{\beq}{\begin{equation}}
\newcommand{\eeq}{\end{equation}}

\newcommand{\Msun}{$ M_\odot$}

\newcommand{\mjypb}{mJy~beam$^{-1}$}

\newcommand{\njp}{New J. Phys.} 
\newcommand{\araa}{Annu. Rev. Astron. Astrophys.}   
\newcommand{\apj}{Astrophys. J.}   
\newcommand{\apjs}{Astrophys. J. Suppl. Ser.}   
\newcommand{\aap}{Astron. Astrophys.}   
\newcommand{\aapr}{Astron. Astrophys. Rev.}   
\newcommand{\mnras}{Mon. Not. R. Astron. Soc.}   
\newcommand{\nastro}{Nat. Astron.} 

\title{Cold gas removal from the centre of a galaxy by a low-luminosity jet}

\begin{document}

\pagenumbering{gobble} 

\author{Suma Murthy$^{1,2,3}$, Raffaella Morganti$^{2,1}$,  Alexander Y. Wagner$^4$, Tom Oosterloo$^{2,1}$, Pierre Guillard$^{5,6}$, Dipanjan Mukherjee$^7$ \& Geoffrey Bicknell$^8$}

\maketitle

\begin{affiliations}
 \item Kapteyn Astronomical Institute, University of Groningen, Landleven 12, 9747 AD Groningen, The Netherlands
 \item ASTRON, the Netherlands Institute for Radio Astronomy, Oude Hoogeveensedijk 4, 7991 PD Dwingeloo, The Netherlands.
  \item Joint Institute for VLBI ERIC, Oude Hoogeveensedijk 4, 7991 PD Dwingeloo, The Netherlands.
 \item Center for Computational Sciences, University of Tsukuba, 1-1-1 Tennodai, Tsukuba, Ibaraki, 305-8577
 
  \item Sorbonne Universit\'e, CNRS, UMR 7095, Institut d’Astrophysique de Paris, 98 bis bd Arago, 75014 Paris, France
 \item  Institut Universitaire de France, Minist\`ere de l’Enseignement Sup\'erieur et de la Recherche, 1 rue Descartes, 75231 Paris Cedex F-05, France
 
 \item Inter-University Centre for Astronomy and Astrophysics, Post Bag 4, Pune - 411007, India
 \item Australian National University, Research School of Astronomy and Astrophysics, Cotter Rd., Weston, ACT 2611, Australia
\end{affiliations}

\begin{abstract}
The energy emitted by active galactic nuclei (AGN) may provide a self-regulating process (AGN feedback) that shapes the evolution of galaxies. This is believed to operate along two modes: on galactic scales by clearing the interstellar medium via outflows, and on circum- galactic scales by preventing the cooling and accretion of gas onto the host galaxy. Radio jets associated with radiatively-inefficient AGN are known to contribute to the latter mode of feedback. However, such jets could also play a role on circum-nuclear and galactic scales, blurring the distinction between the two modes. We have discovered a spatially-resolved, massive molecular outflow, carrying \apx75\% of the gas in the central region of the host galaxy of a radiatively-inefficient AGN. The outflow coincides with the radio jet 540 pc offset from the core, unambiguously pointing to the jet as the driver of this phenomenon. The modest luminosity of the radio source ($L\rm_{1.4 GHz}=2.1 \times 10\rm^{23}~\rm W~\rm Hz^{-1}$) confirms predictions of simulations that jets of low-luminosity radio sources carry enough power to drive such outflows. Including kpc-scale feedback from such sources -- comprising of the majority of the radio AGN population -- in cosmological simulations may assist in resolving some of their limitations.
\label{Abstract}
\end{abstract}

The contribution of various classes of active galactic nuclei (AGN) to feedback and the scales at which they operate are yet to be well characterised. In this context, the models of galaxy evolution only consider large, powerful radio AGN that are capable of heating the intergalactic medium and thereby preventing the gas from accreting on to the host galaxy\cite{McNamara2012}.

However, recently various studies have found that radio AGN also have an impact on galactic scales where the radio jets interact strongly with the interstellar medium (ISM) and drive multi-phase outflows, of which dense cold gas is the most massive component\cite{Veilleux2020}. Numerical simulations predict that such  jet-ISM interactions could also be strong in low-luminosity radio AGN where the jets spend more time embedded within the host galaxy\cite{Sutherland2007, Wagner2012,Mukherjee2016,Mukherjee18b}. Observationally, to understand the impact of radio jets on the ISM, we need to spatially resolve the site of jet-ISM interaction and also be able to disentangle the contribution of the radiation from the optical AGN and of the radio jets to the observed impact on the cold gas. Such studies have so far not been possible. We present the case of a low-luminosity radio AGN where spatially resolved molecular gas observations have shown the presence of a massive outflow that is entirely driven by the radio jets. 

The source under consideration, \target, is a radio galaxy which consists of a bright, kpc-scale structure and large, low-surface brightness radio lobes \apx240 kpc in size. The kpc-scale emission represents the current phase of activity which started only between 400 thousand and 900 thousand years ago\cite{Brienza2018, Giroletti2005}.  
The central radio source is nested in  NGC~1167 ($z= 0.0165$\footnote{We assume a flat Universe with $H_0$ = 67.3 \kmps\ Mpc\p{1}, $\rm\Omega_{\Lambda}$ = 0.685, and $\rm \Omega_{M}$ = 0.315\cite{Collaboration2014PlanckResults} for all our calculations. At $z=0.0165$, 1$''$ corresponds to 0.349 kpc.}), a gas-rich, massive early-type galaxy and the optical AGN in NGC~1167 is found to be radiatively inefficient\cite{Ho2009}. It has been suggested that the cold gas in the inner few kiloparsecs of the galaxy shows signatures of disturbed kinematics\cite{Murthy19}. Furthermore, the diffuse large-scale lobes are caused by a previous phase of activity and indicate that \target\ is a radio source which has undergone multiple episodes of activity\cite{Shulevski12,Brienza2018}. Thus, its properties make \target\ an important case for tracing the interaction between the radio jets and the ambient gas. We use the cold molecular gas as a tracer for this purpose because it has been found to be typically the most massive component (i.e. more massive than the warm ionised gas) of AGN-driven outflows.

\section{A massive molecular outflow} \label{sec:outflow}

To probe the distribution and kinematics of the cold molecular gas in the nuclear region of \target, we carried out CO(1-0) observations with NOrthern Extended Millimeter Array (NOEMA) at an angular resolution of $1.9'' \times 1.5''$. At the redshift of the source, this corresponds to a spatial extent of 540 pc. We detect a circumnuclear molecular gas structure, \apx3 kpc in size and a quiescent CO ring, \apx 10 kpc in radius (see Fig.\  \ref{fig:CO_emission}). Figure \ref{fig:PV} shows that the kinematics of the ring is consistent with the regular rotation of the galaxy\cite{Struve10c} (see Methods). The kinematics of the gas in the central few kiloparsec, however, entirely deviates from this large-scale regular rotation.

The kinematics of the circumnuclear gas are of particular interest since this gas is spatially coincident with the radio jets.  As can be seen from the velocity dispersion map (Fig.\ \ref{fig:CO_emission} bottom-right) and the position-velocity plots of Fig.\ \ref{fig:PV}, the gas barely shows sign of regular rotation, suggesting that any disc structure that may have existed has been mostly disrupted. The velocity dispersion presented in Fig.\  \ref{fig:CO_emission} shows that the gas is particularly disturbed in the region 540 pc south-east of the radio core where the southern radio jet undergoes a sharp bend. At this location,  the  gas exhibits a particularly high velocity dispersion (with a maximum value of \apx250 \kmps) and is outflowing at large velocities, blueshifted up to \apx500 \kmps\ with respect to the systemic velocity (see Fig.\  \ref{fig:PV}). Furthermore, the CO emission is also significantly brighter in this region compared to the rest of the circumnuclear structure (see Fig.\ \ref{fig:CO_emission}, top-right). This CO `hotspot', corresponding to the location of the outflowing gas, suggests that the gas in this region may have a different (higher) excitation temperature or originates from optically thin gas as a result of the mechanism that is producing the outflow\cite{Oosterloo2017,Oosterloo2019}. These findings, especially the spatial offset from the radio core, strongly suggest a link between the outflow and the radio jet instead of the outflow being driven by the nucleus itself. 

Depending on the choice of CO to H$_2$ conversion factor, the mass of the molecular gas ($M_{H_2}$) in the entire circumnuclear structure ranges between  $(6.7 \pm 0.7) \times 10^{6}$ \Msun\ and $(15.7 \pm 1.6) \times 10^{6}$ \Msun\ (see Methods for mass and outflow rate estimates). The mass of the gas that is disturbed by the interaction ranges between  $(5.0 \pm 0.7) \times 10^{6}$ \Msun\  and $(11.7 \pm 1.6) \times 10^{6}$ \Msun. Thus, about 75\% of the emission arising from this region is associated with the outflowing component. We derive a mass outflow rate ranging between 5 \Msun\ yr\p{1} and 10 \Msun\ yr\p{1}. This implies that within the short life-span of the radio jet, lasting only for a few million years, the kpc-scale molecular gas reservoir will be entirely depleted. The escape velocity of the galaxy estimated based on the \hi\ rotation curve\cite{Struve10c} is \apx500 \kmps, higher than the outflow velocities observed. Thus the gas removed from the nuclear region will rain down onto the galaxy at a later time.

\section{Radio jet as the trigger of the molecular outflow} 

The location and the properties of the molecular-gas outflow suggest that the radio jet is responsible for the massive outflow observed in \target. The energetics of the phenomena involved support this scenario.

The estimated kinetic power of the gas associated with the molecular gas outflow\cite{Holt2006} ranges between 1.8 $\times$ 10\pp{41} erg s\p{1} and 1.9 $\times$ 10\pp{42} erg s\p{1}. NGC~1167 has been classified as a low-luminosity optical AGN (LLAGN) with a spectrum typical of low-ionisation nuclear emission-line region (LINER) galaxies\cite{Ho2009}. 
Estimates of the bolometric luminosity of the AGN range between 3~$\times$~10\pp{41}~erg~s\p{1} and 7~$\times$~10\pp{42}~erg~s\p{1} (see Methods for details on these estimates). Thus, radiation can drive the outflow only if the bolometric luminosity is at the higher end of the estimated range and the optical AGN transfers energy to the ISM with a very high efficiency. This, combined with the  outflow being offset from the nucleus  at the location where the jet is bent, makes it unlikely that radiation can explain the observed outflow and its properties. 

On the other hand, estimates of the radio-jet power\cite{Willott99, Cavagnolo10} range between $8.2\times 10$\pp{43} erg s\p{1} and $1.3\times10$\pp{44} erg s\p{1} (see Methods), about two orders of magnitude higher than the gas kinetic power. We note that the estimate of the jet power is based on various correlations (see Methods) that may underestimate the actual jet power for low-luminosity radio sources\cite{Mukherjee18a}. As such, the estimate obtained here should be regarded as a lower limit. Nevertheless, this shows that the radio jet can drive the observed outflow even at a low efficiency.


The possibility of low-luminosity radio jets impacting the surrounding medium has been suggested by earlier studies, which however have limitations compared to the results presented here. Those studies have focussed on Seyfert galaxies, low-$z$ quasars with low-luminosity radio jets\cite{Capetti1996,Wilson1999,Morganti15,Garcia-Burillo2014,Venturi2021,Jarvis2019,Husemann2019}, and on a number of LLAGN\cite{Alatalo11,Combes2013,Riffel2014, Rodriguez-Ardila2017, Fabbiano18b}. In the group of radiatively efficient AGN (Seyferts and quasars), the effect of the nuclear wind and radiation, and the effect of the radio jets are difficult to disentangle and hence it is not possible to quantify the role and impact of the latter. In the case of LLAGN, the impact of the jet is only indirectly inferred as a consequence of the low bolometric luminosity and hence the inability of the radiation/wind to drive the outflows; however, there has been no direct evidence to date showing that the radio jets are indeed the cause of such outflows. Moreover, a majority of these cases only show the presence of less massive warm-ionised gas outflows which are not significant in the context of feedback. In a small subset where molecular gas outflows do exist, the outflow has not been localised with respect to the radio jet.

The best example of the impact of a low-luminosity radio jet on the molecular gas has been observed in the Seyfert 2 galaxy IC\,5063\cite{Morganti15,Oosterloo2017,Mukherjee18a}. Here the jet-ISM interaction has been `caught in action' by the high-spatial resolution observations that enabled the localisation of the cold-gas outflow which showed that radio jets could be the main driver of the outflow. However, in this case, the optical AGN is powerful and hence it is not possible to completely exclude the effect of the strong nuclear radiation on the outflow.

Thus, \target\ is the first clearest case where a young, low-luminosity radio jet is unambiguously found to be responsible for driving a massive molecular gas outflow which has also been spatially localised. Furthermore,  detailed optical integral field unit (IFU) mapping of the host galaxy shows that there is no ionised gas outflow in \target\cite{Gomes2016WarmGalaxies}. Thus, we confirm, as has been found in a number of other objects\cite{Veilleux2020}, that also in this case, the cold molecular gas is the most massive component of the outflow.


\section{Comparison with a hydrodynamic jet-ISM simulation} \label{sec:simcomp}



The results presented here are of significance to theoretical models of jet-ISM interactions in that they confirm some of their predictions. Numerical simulations\cite{Sutherland2007, Wagner2012, Mukherjee2016, Mukherjee2017, Mukherjee18a, Mukherjee18b} have shown that radio jets, despite being collimated structures, can impact the host galaxy  significantly over a large volume when expanding into a clumpy ISM. Broadly speaking, they predict that (i) this impact is large in the early phase of their life; (ii) low-luminosity jets remain trapped in the ISM while trying to break through the gas, continuously injecting their energy into the ISM and their impact over time becomes very pronounced. Our results show that this indeed is the case in \target. 

To explore this further, we compare our observations with a relativistic hydrodynamical simulation of jet-ISM interactions from Mukherjee et al\cite{Mukherjee18b}. The simulation is not a tailored simulation for \target, so this comparison is meant to qualitatively illuminate the underlying physics rather than represent a quantitatively accurate model of the source. The particular simulation we chose for the purpose is Simulation D, in which a relativistic jet of power $P_\mathrm{jet}=10^{45}$ erg s\p{1} propagates through a thick galactic disc with a clumpy gas distribution at a tilt angle of $45^\circ$, a choice based on the results from \hi\ absorption studies\cite{Murthy19} which indicate that the radio jets are very likely in the process of expanding into a gaseous disc.  

A jet with an order of magnitude lower power would be at the lower limit of being capable of generating the outflow seen in \target, since velocities reached by clouds when dispersed in energy-driven jet bubbles typically scale as $P_\mathrm{jet}^{0.2}$\cite{Wagner2011RelativisticGalaxies}. The jet power implied by the jet-driven ISM dynamics is between $10^{44}$ erg s\p{1} and $10^{45}$ erg s\p{1}, and is therefore approximately an order of magnitude larger than the jet power obtained from radio-power scaling relations.

Figure~\ref{fig:simslices} shows a series of mid-plane density slices perpendicular to the $y$-axis of the simulation that highlight the evolution of the jet-disc system. The lower-density jet-plasma of the backflow in the cocoon, secondary jet streams within the disc, and the bow-shock bounded bubble are also clearly visible. The disc in the simulation is 4 kpc in diameter and by 0.2 Myr the jet has processed the ISM in the central 2 kpc. The jet-ISM interactions are quite complicated and most of the gas in the central region is strongly dispersed by the jet. The dispersion of dense gas is the strongest during the first few 100 kyr, and by 0.8 Myr, a substantial amount of jet plasma is venting through chimneys perpendicular to the disc reducing the energy coupling between jet and gas. The main jet streams interact directly with clumps in their path and become deflected or split, brightening in radio emission as a result of shocks, while generating strong gas velocity dispersions and outflows. 

Intriguingly, we see a similar effect in \target, where the massive outflow and a strong deflection in the radio continuum are located co-spatially along the southern jet. Moreover, we also do not see a disturbance in the gas along the much weaker counter jet. As the simulation shows, asymmetric signatures of jet-ISM interactions are expected if the ISM is clumpy\cite{Gaibler2011AsymmetriesInteraction}. The northern jet stream may be propagating through a diffuse inter-cloud medium while the southern jet may be directly hitting a molecular cloud resulting in the observed radio morphology, a feature often seen in radio sources, for example, IC\,5063\cite{Morganti15}. The density slices also show a largely evacuated central 0.5 kpc region that was cleared by the jet, a feature reminiscent of the apparent eradication of the inner kpc disc of \target\ in the observations. 



We looked for signatures of high-velocity bulk outflows at different time snapshots of the simulation and at different lines-of-sight through the simulation box to compare with the observations (see Fig~\ref{fig:pv-diagrams} and Methods for more details). We find that regions of jet-ISM interactions show enhanced velocity dispersions of the dense phase (gas with densities $n>100$ cm$^{-3}$), and that clouds hit directly by the main jet streams may be accelerated to beyond 500 \kmps, consistent with our observations. The signatures of cloud acceleration from our simulations are also offset from the centre. These dense gas dispersion and outflow signatures are the strongest at around 200 kyr since the start of the jet activity, and drop as jet plasma gradually leaks out of the galaxy. The simulation suggests, therefore, that the inner jet in \target\ driving the outflow shown in Fig.~\ref{fig:PV} is likely younger than a Myr, in line with the estimated age of 0.9 Myr or less by various observational studies\cite{Giroletti2005, Brienza2018}. Bulk outflow velocities are also higher the closer the jet is aligned to the line of sight. Together with the imposed $38^{\circ}$ inclination of the disc to the line-of-sight, a simulation similar to that used in the comparison study here, but fully tailored in its initial conditions to \target, can constrain the three-dimensional orientation of the jet.


\section{Relevance of radio jets for feedback}

The results on B2~0258+35 are relevant in the broader context of AGN feedback. Radio-loud AGN are commonly found in massive galaxies like NGC\,1167 (i.e.\ ${M_*} \geq 10^{10.5}$\Msun). It has been shown that 30\% of  massive galaxies host radio sources with luminosities less than $ 10^{23}$ W Hz$^{-1}$, compared to less than 1\% of the massive galaxies hosting powerful radio sources ($\log L_{\rm 1.4~GHz} \geq 25$ W Hz$^{-1}$)\cite{Best2005,Sabater2019}. Furthermore, various studies have also shown that cold gas is more commonly present in the nuclear region of young radio sources\cite{Morganti2018TheAbsorption}. Thus, our results highlight that the galactic-scale impact of low-power radio galaxies may represent an important component -- so far largely neglected -- for models of AGN feedback, provided the radio emission can efficiently couple with the surrounding medium, which our results suggest to be the case.

This makes these AGN (and their impact) relevant for cosmological simulations. \target\ is a restarted radio galaxy with a short time gap of a few tens of Myr between the dimming of one phase of radio emission and the starting of the new episode\cite{Brienza2018}. Our results illustrate that even in the renewed phase of activity, the radio jet is able to impact the host galaxy significantly. This recurrent impact of the AGN on the host galaxy over multiple cycles is  one of the requirements of cosmological simulations to explain the observations\cite{Genel2014IntroducingTime, Sijacki2015TheTime,Schaye2015TheEnvironments, Crain2015TheVariations}.

The inclusion of kpc-scale feedback from low-luminosity radio sources may also help resolve some of the tensions that exist between observations and simulations\cite{Weinberger17}. At the moment, cosmological simulations have neither the spatial resolution nor the dynamic range in density to capture kpc-scale outflows in detail. However, the first steps in including these features are being taken\cite{Dubois12, Dubois2013, Talbot2021, Talbot2021Blandford-ZnajekJets} and they highlight the significance of jet-driven outflows. In the future, as computational power increases and AGN feedback will be modelled in more detail with zoom-in simulations, observations of multi-phase outflows, and in particular observations of cold molecular outflows similar to our study, will provide invaluable constraints on parameters such as the amount of gas affected, interaction timescales, energy deposition rate, gas redistribution and turbulence generated.



\begin{addendum}
  \item[Data availability] The data presented in this paper were observed with NOEMA under the project ID S20BH. The reduced data products are available here: \hyperlink{https://astrodrive.astro.rug.nl/index.php/s/e3RTJ5wnpGFGFd9}{https://astrodrive.astro.rug.nl/index.php/s/e3RTJ5wnpGFGFd9} or from the corresponding authors on reasonable request. The simulation data for which results are shown in our paper are results from the simulations of Mukherjee et al\cite{Mukherjee18b} and available in raw form:\\ \hyperlink{https://www2.ccs.tsukuba.ac.jp/Astro/Members/ayw/depot/B2_0258+35/data/}{https://www2.ccs.tsukuba.ac.jp/Astro/Members/ayw/depot/B2\_0258+35/data/} where data.0000.flt, data.0037.flt, data.0100.flt  are the data for time snapshots at t = 0, t = 0.2 Myr, and t = 0.8 Myr, respectively, shown in the paper in Figures 3, 4, and 5. These data are in a standard single precision flt binary output format of the hydrodynamic code PLUTO. that combines all thermodynamic variables into one file for one snapshot in time.
  
  \item[Code availability] The data were reduced using publicly available softwares GILDAS and AIPS. The scripts with which the data cubes from the simulations were analyzed are a part of the following repository:\\ \hyperlink{https://bitbucket.org/pandante/pluto-ug-simulation-analyzer/}{https://bitbucket.org/pandante/pluto-ug-simulation-analyzer/}. In particular, the branch B2\_0258+35 contains the specific tools used to obtain the analysis results and plots presented in Figures 3, 4, and 5. The simulations in Mukherjee et al\cite{Mukherjee18b} were performed with the PLUTO code, publicly available here: \hyperlink{http://plutocode.ph.unito.it/}{http://plutocode.ph.unito.it/}. A version of the code that includes routines to set up the simulations is available here:\\ \hyperlink{https://github.com/aywander/pluto-outflows.git}{https://github.com/aywander/pluto-outflows.git} and requires additional setup data files available here:\\ \hyperlink{https://www2.ccs.tsukuba.ac.jp/Astro/Members/ayw/depot/P45_dir45/setup/}{https://www2.ccs.tsukuba.ac.jp/Astro/Members/ayw/depot/P45\_dir45/setup/}. The simulation also requires initial data of the clumpy fractal multiphase gas generated by the code pyFC, also publicly available here:\\ \hyperlink{https://pypi.org/project/pyFC/}{https://pypi.org/project/pyFC/}.

 \item This work is based on the observations carried out under project number S20BH with the IRAM NOEMA Interferometer. IRAM is supported by INSU/CNRS (France), MPG (Germany) and IGN (Spain). We thank the IRAM staff for making these observations possible. SM thanks Orsolya Feher for the help with data reduction. We thank Giuseppina Fabbiano and Mislav Balokovi\'{c} for sharing the results from the \textit{Chandra/NuSTAR} observations prior to publication and Marcello Giroletti for sharing the FITS image of the 8.5 GHz radio continuum emission. The simulations in this paper were undertaken with the assistance of resources from the National Computational Infrastructure (NCI Australia), an NCRIS enabled capability supported by the Australian Government. AYW is supported by JSPS KAKENHI Grant Number 19K03862.
 This preprint has not undergone post-submission improvements or corrections. The Version of Record of this article is published in \textit{Nature Astronomy}, and is available online at \url{https://doi.org/10.1038/s41550-021-01596-6}
 
  \item[Author contributions] SM and RM conceived the project. SM, RM, PG and TO wrote the observing proposal. SM reduced the data. SM, RM and TO carried out the analysis. AYW, DM and GB contributed to the simulations. SM, RM, and AYW wrote the manuscript. All the authors discussed the results and commented on the manuscript.
  
  \item[Competing Interests] The authors declare that they have no competing financial interests.

 \item[Correspondence] Correspondence and requests for materials
should be addressed to Suma Murthy~(email: murthy@jive.eu) or Raffaella Morganti~(email: morganti@astron.nl).
\end{addendum}

\clearpage

\begin{figure}
\centering
\includegraphics[width=\linewidth]{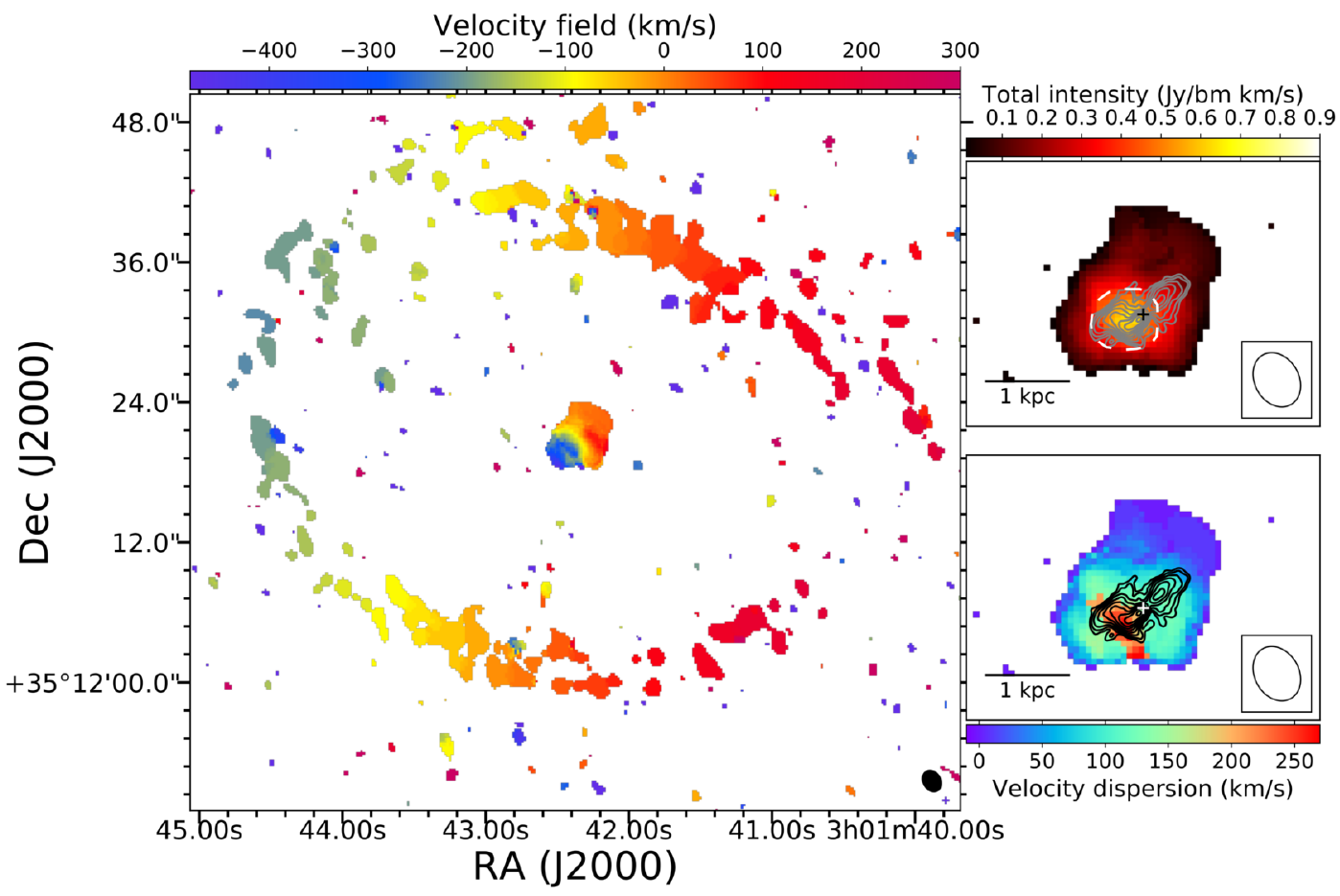}
\caption{{\bf CO emission from NGC~1167}. The left panel shows the distribution and the velocity field of the large-scale ($\sim$10 kpc radius) ring and the inner nuclear region. The kinematics of the ring is regular and consistent with the large \hi\ disc\cite{Struve10c}. The panels on the right show the zoom-in of the nuclear region (i.e.\ the inner few kpc). The top-right panel shows the total intensity map and the bottom-right panel shows the velocity dispersion map. The beam is shown in the bottom right corner of the images and has a size of $1.9'' \times 1.5''$ with a position angle of 29.2$^\circ$. The black contours represent the radio continuum emission at 8.5 GHz\cite{Giroletti2005}. The radio core is marked with a cross. The brightest CO emission and the highest velocity dispersion are observed in a region offset to the south-east from the radio core where the southern radio jet bends strongly. The region considered to estimate the mass of the kinematically disturbed outflowing gas (also see Fig.\ \ref{fig:PV}) is marked by the white dashed lines in the total intensity map (top right).}
\label{fig:CO_emission}
\end{figure}

\begin{figure}
\centering
\includegraphics[width=\linewidth]{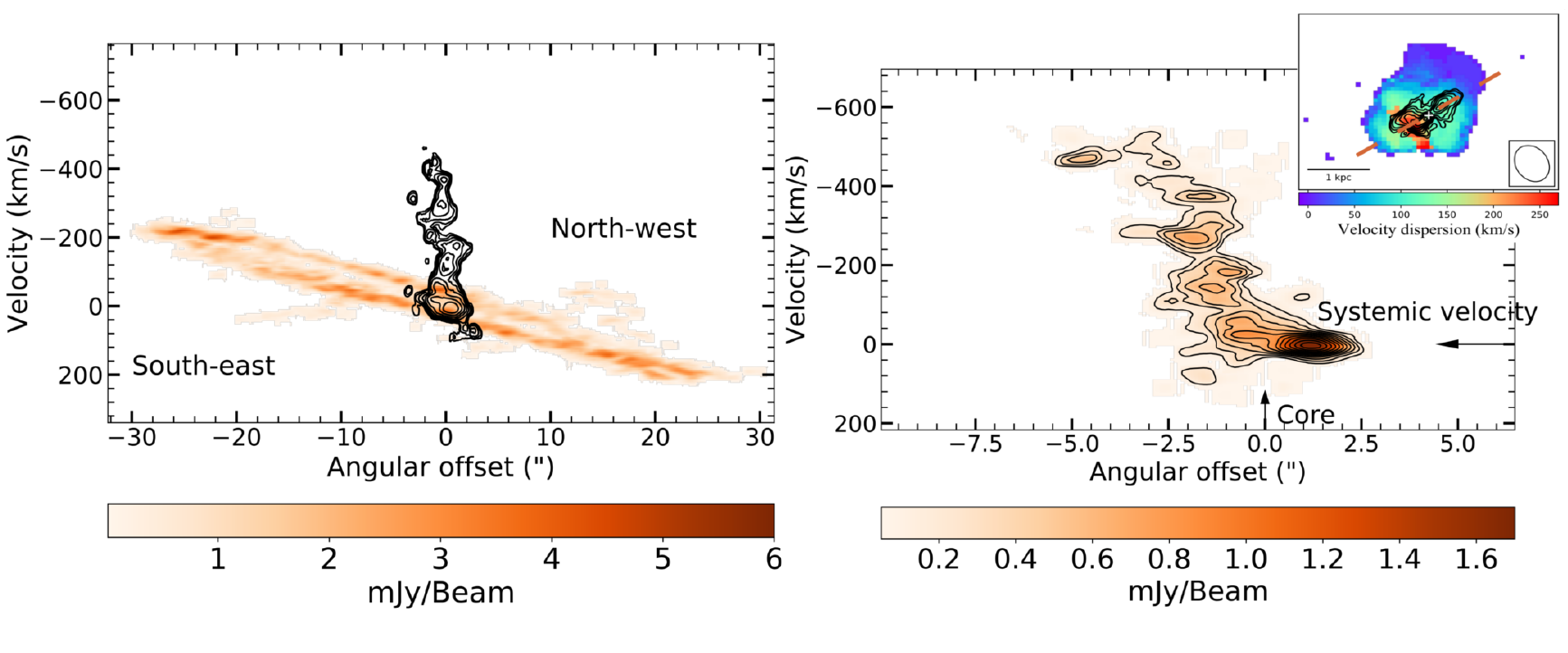}
\caption{{\bf Kinematics of the cold gas}. Left panel shows the position-velocity (PV) plot of the large-scale gas disc in colour scale and that of the circumnuclear gas in contours, both extracted along the major axis of the large-scale gas disc. This plot highlights the extreme difference in the kinematics of the regularly rotating disc and the disturbed gas in the nuclear region. The right panel shows the PV diagram of the circumnuclear gas extracted along the radio axis as shown in the right-panel inset. 
The systemic velocity and the radio core are indicated in the right panel. The kinematics of the gas  distinctly deviates from regular rotation and the outflow is offset to the south-east of the radio core (see also Fig. \ref{fig:CO_emission}).}
\label{fig:PV}
\end{figure}



\begin{figure}
    \centering
    \includegraphics[angle=0, width=0.75\textwidth]{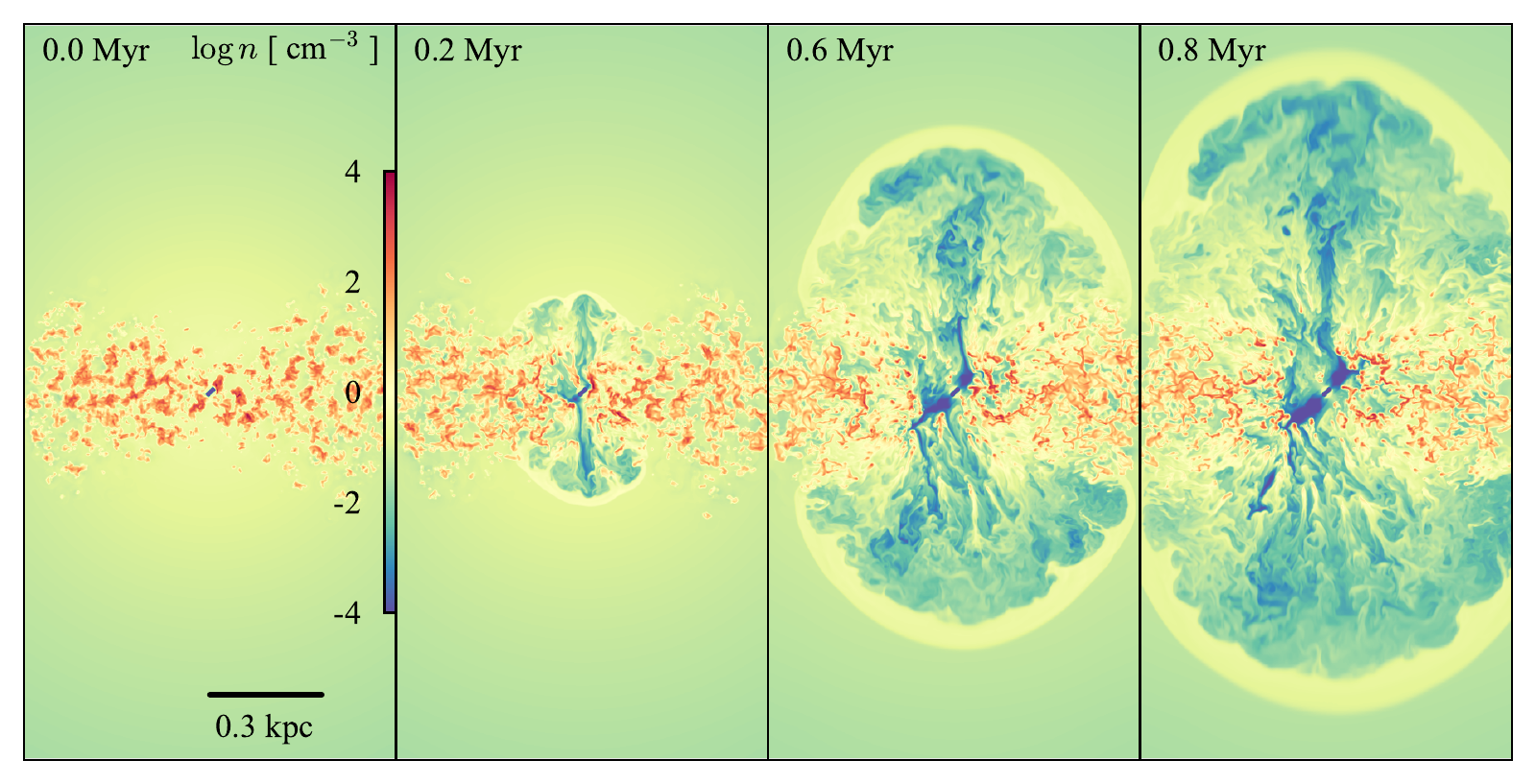}
    \caption{\textbf{Snapshots from the simulations.} Mid-plane logarithmic density slices of simulation D from Mukherjee et al.\cite{Mukherjee18b} at $y=0$ showing the evolution of the jet-disc system. The jet plasma is in blue, the dense clouds in orange-red. Ablated gas and shocked ambient medium is in yellow. The strongest interactions occur within the disc where the main jet stream hits clouds head-on. These regions show enhanced velocity dispersions and bulk velocities up to 500 \kmps\ (see PV diagrams in Methods), and are location of sharp jet deflection and splitting. While the outer disc is dispersed but remains largely intact, the inner 0.5 kpc region is largely cleared of gas by the jet by $\sim 1$ Myr.}
    \label{fig:simslices}
\end{figure}

\begin{figure}
\centering
\includegraphics[width=\linewidth]{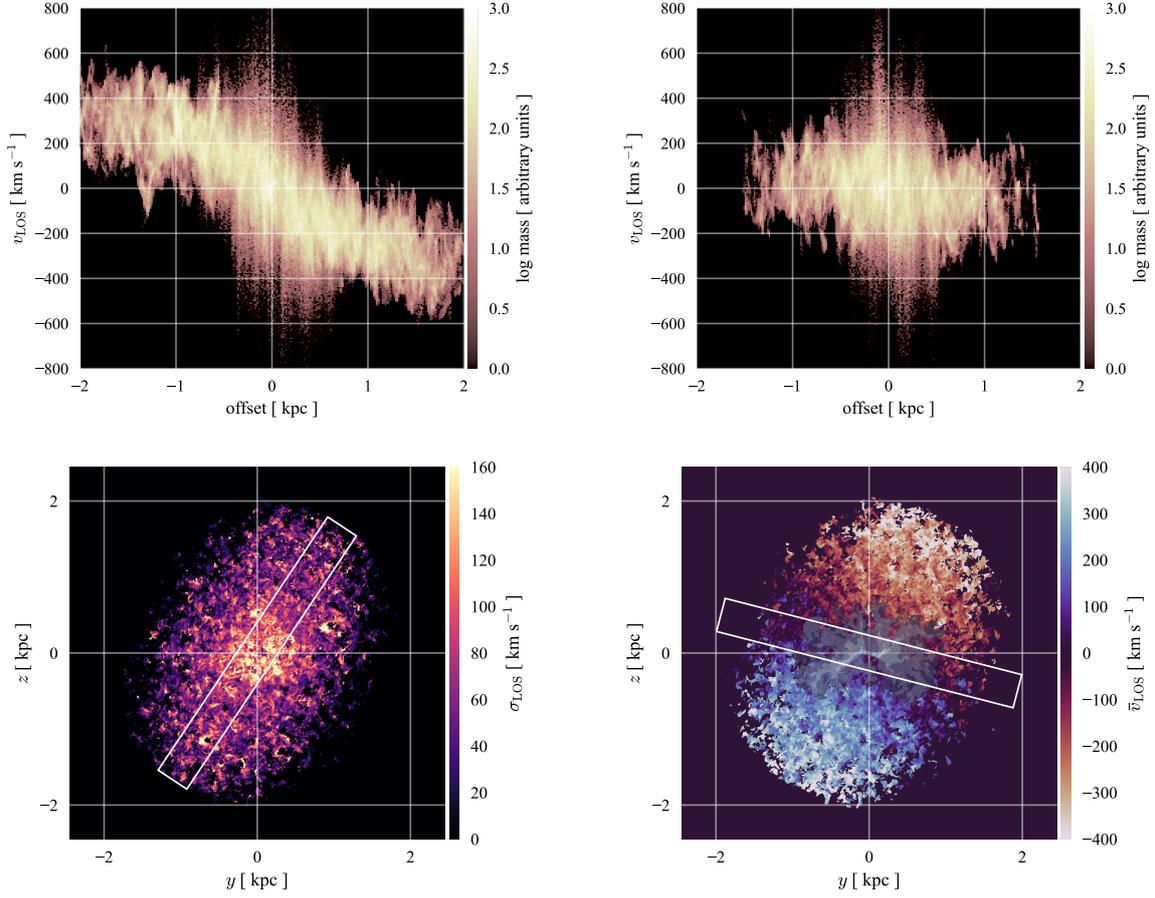}
\caption{\textbf{Synthetic position-velocity diagrams.} 
Top row: PV diagrams of the dense gas (number density greater than 100 cm$^{-3}$) in simulation D from Mukherjee et al.\cite{Mukherjee18b} along the line-of-sight and the slits shown in the corresponding panels of the bottom row. The jet-cloud interactions at $0.2$ Myr produce signatures of enhanced velocity dispersions with some clumps accelerated to beyond 500 \kmps, seen as strong spiky features in the PV diagrams.
Bottom row: Left panel shows the line-of-sight velocity dispersion (second moment) of dense gas ((number density greater than 100 cm$^{-3}$). Right panel: the mean velocity (first moment) showing the rotating disc; the broad-band radio emission is shown in grey scale. The emissivity of the jet plasma is assumed to be proportional to $p^{1.8}$, where $p$ is the plasma pressure\cite{Wagner2011RelativisticGalaxies}.}
\label{fig:pv-diagrams}
\end{figure}

\clearpage

\section*{Methods}


\subsection{\underline{Target information.}} B2 0258+35 is a  low-luminosity radio AGN ($L_{1.4GHz} = 2.1 \times 10^{23}$ W Hz$^{-1}$). It consists of a compact-steep spectrum (CSS) source, about 1 kpc in size\cite{Giroletti2005}, and diffuse 240 kpc low-surface brightness radio lobes\cite{Shulevski12}. The radio spectrum of the central CSS source peaks at 70 MHz, typical of sources belonging to this class. The kpc-sized inner source consists of a core and asymmetric jets (see Fig.\  \ref{fig:CO_emission}). The radio emission is the brightest in the region of where this jet is bent. The age of the central source is estimated to range between 0.4 Myr to 0.9 Myr\cite{Brienza2018}. The large-scale low-surface brightness radio lobes have been estimated to be about 110 Myr old\cite{Brienza2018}. A detailed spectral-index study suggests that they are not old remnants but instead are still being fuelled at a very low rate, perhaps from some `leakage' form the central source\cite{Murthy19}. It has also been suggested that the duty cycle of this AGN is short with a time gap between the two episodes of activity no more than a few tens of Myr\cite{Brienza2018}.


The host galaxy, NGC~1167, is a gas-rich ($M_{\rm \hi} = 1.5 \times 10^{10}$ \Msun) early-type galaxy, with a 160 kpc diameter regularly-rotating \hi\ disc. The disc has very regular kinematics within a radius of 65 kpc, and shows signs of interactions, perhaps with a satellite gaalxy, only in the very outer parts. This suggests that the galaxy has not undergone a major merger in the last few billion years\cite{Struve10c}. CO(1-0) emission from the central 8 kpc of the galaxy has been detected earlier using single-dish observations\cite{OSullivan2015} and very faint CO emission from a region outside the central 8 kpc was detected using interferometric observations of poorer sensitivity than the single-dish studies\cite{Bolatto2017}. Our observations at higher spatial resolution and sensitivity show that the CO emission at large scales arises from a large ring of molecular gas (see Fig.\ 1) of \apx10 kpc radius. This ring shows regular rotation consistent with that of the large \hi\ disc and overlaps with low-level star-formation activity along faint spiral arms-like structures\cite{Gomes2016Spiral-likeGalaxies}. 

\subsection{\underline{Observations and data reduction.}}
The NOEMA observations of CO(1-0) in \target\ (Proposal ID: S20BH) were carried out over five observing runs in October and November, 2020 with the telescope in C configuration with either nine or ten antennas. The setup included the lower and the upper sidebands centred at 96.7 GHz and 112.24 GHz respectively, each of bandwidth of bandwidth 7.72 GHz subdivided into 3859 channels, giving us a spectral resolution of 2 MHz or 5.2 \kmps. 3C84 was used as the bandpass calibrator while 2010+723, LKHA101, MWC349, 0420-014 and 081+202 were the flux calibrators used over different observing runs and J0304+338 was used for phase and amplitude calibration. We observed the target source, \target, for 10.5 hours in total. 

We carried out calibration, flagging of bad visibilities, and averaging the two polarisations using standard pipelines in Grenoble Image and Line Data Analysis Software (GILDAS). Then we exported the calibrated \textit{uv} tables in uvfits format for further reduction and analysis in Astronomical Image
Processing Software (AIPS)\cite{Greisen2003AIPSVLBA}. Since our primary focus is on the CO(1-0) emission, we further reduced only the data from the upper side band which covered the frequencies of interest. We first self-calibrated the data: initially a few cycles of phase-only self-calibration followed by a round of amplitude and phase self-calibration. Then we subtracted a first order polynomial from each calibrated visibility spectrum to obtain the continuum-subtracted \textit{uv} data.

To improve the signal-to-noise ratio, we Hanning-smoothed the data and averaged 2 channels together and imaged the continuum-subtracted \textit{uv} data to obtain the spectral cube. We made the cube with natural weighting to obtain maximum sensitivity. The cube has an angular resolution of 1.93$''$ $\times$ 1.5$''$ with a position angle of 29.2$^\circ$, a spectral resolution of 42 \kmps, an RMS noise of 0.4 \mjypb\ per 42 \kmps\ channel. We extracted the moment maps from this cube by using a mask to include emission above 3-$\sigma$ in the line-channels.


\subsection{\underline{Molecular gas mass, mass outflow rate, kinetic power.}}

We detect a large-scale CO ring as well as a circumnuclear structure. We estimated the molecular gas mass of the entire circumnuclear gas, the outflow i.e., the gas mass corresponding to the CO hotspot (see Fig.\ 1) and also the CO ring. This was done using the relation $M_{\rm H_2} = \alpha L_{\rm CO}$  where $M_{\rm H_2}$ is the molecular mass in solar mass units, $L_{\rm CO}$ is the CO line luminosity and $\alpha$ is the conversion factor. We estimated the CO line luminosity using the standard relation $L_{\rm CO} = 3.25 \times 10^7$ $S_{\rm CO}$ $\Delta v$ $\nu_{obs}^{-2}$ $D_L^2 (1+z)^{-3}$ K \kmps pc\pp{2} where $S_{\rm CO}$ $\Delta v$ is the velocity integrated CO line flux \cite{Solomon2005}. To estimate the molecular gas masses, we first extracted the flux from the large ring, the circumnuclear gas and the outflowing component in the circumnuclear gas. For the outflowing component, we extracted the flux from the CO hotspot that overlaps with the southern radio jet, marked in Fig. \ref{fig:CO_emission} top-right panel, because the outflowing gas offset from the radio core as seen in the PV diagram shown (Fig.\  \ref{fig:PV}) arises entirely from this hotspot.

We assumed a range of values for the conversion factor: (i) typical for Milky Way-like galaxies\cite{Daddi2010}, $\alpha =3.4$ \Msun (K \kmps pc\pp{2})$^{-1}$, (ii) for ultra luminous infra-red galaxies (ULIRGs) where the ISM is more turbulent due to star formation activity\cite{Downes1998}, $\alpha =0.8$ \Msun (K \kmps pc\pp{2})$^{-1}$ and (iii) for highly turbulent optically thin gas\cite{Bolatto2013}, $\alpha = 0.34$ \Msun (K \kmps pc\pp{2})$^{-1}$. 
For the regularly rotating large ring, we have used $\alpha =3.4$ \Msun (K \kmps pc\pp{2})$^{-1}$ typical of quiescent gas. For the gas in the circumnuclear region, including the outflow, we have estimated H$_2$ mass using the other two conversion factors more suitable for turbulent gas. The resulting H$_2$ masses are presented in Table \ref{tab:outflow_parameters}. We find that the mass associated with the outflowing component corresponds to 75\% of the CO emission from the circumnuclear region.

We followed the prescription in Harrison et al.\cite{Harrison2018} to estimate the mass outflow rate. Following Fig.\ \ref{fig:PV}, the centroid of the outflowing gas is \apx1.5$''$ offset from the radio core. This corresponds to an offset of \apx 540 kpc from the core. We estimated the outflow velocity using the expression\cite{Rupke2013TheMergers} $v_{\rm out} = v_{\rm shift} + 2\sigma$ where $v_{\rm shift}$, the velocity offset of the broad emission wing of the spectrum with respect to the systemic velocity, is 93 \kmps\ and $\sigma$ of the line is 182 \kmps\ estimated via a Gaussian fit to the blueshifted emission wing. We obtain an outflow velocity of \apx458 \kmps. That gives a timescale of $\tau_{\rm dyn}$ \apx1.1 Myr for the gas to reach its present location from the centre. The mass outflow rate will then be $\dot{M} = M_{\rm CO}/\tau_{\rm dyn}$ \Msun yr~\p{1}. We obtain a mass outflow rate ranging between 4.5 \Msun yr\p{1} to 10.5 \Msun yr\p{1} depending on the choice of $\alpha$ as mentioned before. The range of values for different values of $\alpha$ are tabulated in Table \ref{tab:outflow_parameters}.

We estimated the kinetic power of the outflow following Holt et al.\  2006, using the expression $\dot{E}= 6.34 \times 10^{35}~\dot{M}/2~(v_{\rm out}^2 + FWHM^2/1.85)~$erg~s$^{-1}$ where $\dot{M}$ is the mass outflow rate, $v_{out}$ is the rest-frame outflow velocity and FWHM is the full width at half maximum of the emission profile of the outflow. We measured the FWHM of the outflow to be 430 \kmps\ by fitting a Gaussian profile to the emission spectrum. For the range of mass outflow rates, we obtain the kinetic power in the range 1.8 $\times$ 10\pp{41} erg s\p{1} to 1.9 $\times$ 10\pp{42} erg s\p{1} (see Table \ref{tab:outflow_parameters}).


\subsection{\underline{Bolometric luminosity and radio power estimates.}}

NGC~1167 was studied as part of a sample of radiatively inefficient AGN by Ho et al\cite{Ho1997,Ho2009}. We derived the range of bolometric luminosity in two ways following their estimates using \halpha\cite{Ho1997}\ and the X-ray luminosity derived from \textit{Chandra} observations\cite{Ho2009}. Using their fluxes and the updated distance to NGC~1167\cite{Struve10c}, the H$\alpha$ luminosity of the nuclear region is $\log L_{\rm H\alpha} = 40.25$ erg s$^{-1}$.
This can be converted to bolometric luminosity using the relation:
$L_{\rm bol} = 2.34 \times 10^{44} (L_{\rm H\alpha}/10^{42})^{0.86}$ erg s$^{-1}$;
resulting in $\log L_{bol} = 42.87$ erg s$^{-1}$.

There are various estimates of $2-10$ keV X-ray luminosity ($L_{\rm 2-10 keV}$~erg s\p{1}) of NGC 1167. \textit{XMM-Newton} observations\cite{Akylas2009XMM-NewtonDistribution} provide log ($L_{\rm 2-10 keV}$~erg s\p{1})$=39.6$ while higher spatial resolution \textit{Chandra} observations\cite{Ho2009} provide an estimate of log ($L_{\rm 2-10 keV}$~erg s\p{1})$=40.32$. A study based on ASCA data\cite{Panessa2006OnGalaxies} estimate ($L_{\rm 2-10 keV}$~erg s\p{1})$=42.07$ after correcting for an absorption due to a Compton-thick AGN. However, new \textit{Chandra/NuSTAR} observations clearly show that the AGN is not Compton thick (\textit{Chandra} cycle-22, program 22700176; PI: Fabbiano; \textit{in preparation}) and hence a correction for absorption is not required. Thus we use the X-ray luminosity estimate of ($L_{\rm 2-10 keV}$~erg s\p{1})$=42.07$ obtained with the earlier \textit{Chandra} observations. Adopting the conservative bolometric correction\cite{Ho2009} $C_X = {L}_{\rm bol}/{L}_{\rm X} \approx 15.8$, we obtain a bolometric luminosity of $\log (L_{\rm bol}~\rm erg~\rm s^{-1}) = 41.52$.


The kinetic power of a radio jet is, despite its fundamental importance, a quantity difficult to derive. Relations have been proposed using the radio luminosity as a proxy of the jet power motivated by the results and correlation found with the optical emission lines\cite{Willott99} or the energy required to inflate X-ray cavities \cite{Cavagnolo10}  although their limitations have also been emphasised\cite{Godfrey2016}. For low-luminosity radio jets, it has been found that these relations may actually underestimate the jet power\cite{Mukherjee18a}. Despite the uncertainties, we derive the jet power of \target\ using these scaling relations. Using the relation proposed by Willott et al.\cite{Willott99} 
\begin{math}
{P}_{\rm jet} \approx f^{3/2}~~3 \times 10^{38} ({ L}_{\rm 151\,MHz}/10^{28} \rm{W\, Hz^{-1}\, sr^{-1})^{6/7}}\, \rm W\\
\end{math}
with the luminosity at 150~MHz\cite{Brienza2018} $\log L_{\rm 150\,MHz} = 24.42$ W Hz$^{-1}$ and assuming $f=10$, a parameter taking into account uncertainties like, for example,  fraction of energy in non-radiating particles (see Godfrey \& Shabala\cite{Godfrey2016} for a discussion). In this way we obtain a jet power of ${ P}_{\rm jet} = 43.91$ erg s$^{-1}$.
Using the relation\cite{Cavagnolo10}, $P_{\rm jet}=5.8\times10^{43}(L_{\rm 1.4~GHz}/10^{40})^{0.7}$ erg s$^{-1}$, we derive a jet power of  ${\rm log P}_{\rm jet} = 44.13$ erg s$^{-1}$.

\subsection{\underline{Simulations: properties and synthetic PV diagrams.}} \label{sec:methcomp}

The simulation we used to support the interpretation of our observations is simulation D in Mukherjee et al.\cite{Mukherjee18b}. This is a 3-dimensional grid-based single-fluid simulation whose numerical scheme is capable of treating a large dynamic range in density, temperature, and velocity, and therefore tracing fast hot diffuse outflows, capturing shocks propagating into rapidly cooling media, following dense turbulent gas, and tracing cold outflows. For details on the setup of simulation D see Section 2 in Mukherjee et al.\cite{Mukherjee18b} and their Table 2 for relevant parameters. The simulation employs a jet with power $10^{45}$ erg s$^{-1}$ interacting with the interstellar medium of a gas-rich disc galaxy in a volume of $4 \times 4 \times 8$ kpc$^3$ with a grid resolution of 6~pc. The system consists of radio jets propagating through a thick galactic disc of diameter 4 kpc. The radio jets are tilted at an angle of 45$^\circ$ with respect to the disc, and for the construction of PV diagrams described below, the disc itself is seen at an angle of 38$^\circ$ from the edge-on view. 

Our observations suggest that the gas in the central few kpc of the galaxy is entirely disturbed and the disc, if any, is destroyed to a large extent. Thus, assuming there was a gas disc to begin with, we need to assume its orientation. The host galaxy NGC~1167 has an inclination angle of 38$^\circ$. It can be seen from Fig. \ref{fig:CO_emission} that the kinematics of the gas in the central few kpc is distinct from the large-scale rotation of the galaxy suggesting that it could have a different orientation compared to the large disc. The two stable planes of orientation possible for the circumnuclear disc are: (a) the same as the large-scale disc; (b) perpendicular to the large-scale disc. The latter possibility would imply that the disc is much more edge-on compared to the large disc. However, the observed velocity gradient does not agree with this possibility if a part of the disc is in regular rotation. Thus it is more likely that the circumnuclear disc is along the same plane as the large disc. Hence we chose an inclination angle for the galactic disc that is close to that of NGC\,1167. We further note that the precise angle between the jet and the disc is not known. However, the jet in \target\ is unlikely to be perpendicular to the disc and is likely inclined substantially towards the disc\cite{Murthy19}, and hence we chose a simulation with a jet tilt angle of $45^\circ$ for this comparison.

The estimation of the jet power of \target\ involves uncertainties (see the previous subsection). As deduced for the case of IC 5063\cite{Mukherjee18a}, the jet power may be an order of magnitude higher than what is inferred from the radio power, justifying the choice of a simulation with higher jet power. We emphasize that this simulation is not a precise representation of the jet-disc system of \target\ and a comparison is meant to be illustrative of how jets can generate cold dense outflows which may resemble the observed outflow in \target. 

In the top two panels of Fig.~\ref{fig:pv-diagrams} we show the rotation field of dense gas, above a number density, $n$, of 100 cm$^{-3}$, with temperatures typically in the range of few tens to few 1000 K, together with synthetic broad-band radio emission contours and the two slit orientations used to produce the PV diagrams. The simulation time is $t=0.2$ Myr after jet injection (see second panel in Fig.~\ref{fig:simslices}). These slit placements are equivalent to the region used for the observational data to produce Fig.~\ref{fig:PV}. The rectangular slits are 2 kpc long and 0.45 kpc high. 

In the two panels in the bottom row of Fig.~\ref{fig:pv-diagrams} we show position-velocity diagrams of the dense gas (density $n>100$ cm$^{-3}$) in the simulation extracted from the slits placed along the major axis of the disc or along the jet (see corresponding top-row panels) at time $t=0.2$ Myr since jet injection. Strong gas dispersion along the region impacted by the jet and acceleration of individual clumps are clearly seen as broad spiky features in the PV diagram. 

The features in the PV diagrams obtained from the simulations, including the shape of the rotation curve and the amount and location of dispersion seen, depend strongly on the stage of the jet-ISM interaction (the time-snapshot chosen from the simulations) and on the viewing angle of the system. As with the jet orientation with respect to the disc, the viewing angle with respect to the jet is not well-constrained by the observations, so we inspected over 200 combinations of snapshot times and viewing angle with respect to the jet, for which the disc inclination to the line of sight was $38^{\circ}$. We found that snapshots at approximately 200 kyr since jet injection show the strongest velocity dispersions in regions of jet-ISM interactions, with bulk outflows exceeding 500 \kmps. At later times, for example, at 0.8 Myr (see third panel in Fig.~\ref{fig:simslices}), only weak bulk outflows were seen, as jet plasma increasingly vents through the porous ISM into the galactic halo.



The results by Mukherjee et al.\cite{Mukherjee18b} show that jets that are more closely aligned with the disc exert stronger feedback onto the disc ISM, and the jet may indeed be closer to the plane of the disc than the $45^{\circ}$ in the simulation. The uncertainty in the inclination of jet to disc must always be considered in conjunction with the uncertainty in the inclination of the line-of-sight to the disc and that to the jet. Assuming a disc inclination of $38^{\circ}$ with respect to the line-of-sight, we found that the closer the main jet streams are aligned to the line of sight, the stronger the outflow signatures become. A stronger alignment will, however, also reduce the offset of the outflow signature from the core. From the qualitative comparisons we performed here, we find that the jet is likely substantially inclined toward the line of sight but a simulation tailored to \target\ and a detailed analysis of the radio morphology is required to properly constrain the three-dimensional orientation of the jet and the inclination of the jet relative to the disc.

The synthetic PV diagrams from the simulation display signatures of dispersed gas and accelerated clouds in both the redshifted and blueshifted halves, while the observations only show a one-sided outflow. There are two possible reasons for this discrepancy. One is that, in generating the synthetic PV diagrams, we did not take into account absorption along the line-of-sight. The CO gas is likely optically thick and a redshifted outflow may exist but be obscured by the disc. The other is that the ISM in \target\ is even clumpier than that in the simulations and that the approaching jet encounters a large clump, leading to brightening and deflection of the radio plasma, and while the receding jet is propagating through much lower density media.

Despite the fact that the simulation we used for comparison was not specifically tailored to \target, the strong, off-center outflow features and the timescales they are generated on, the jet deflections, and the clearance of gas in the central regions indicate that some of the physics of the jet-ISM interactions captured in the simulation are indeed operating in \target. 




\clearpage


\begin{table}
\caption{The estimates of molecular gas masses, mass outflow rate and kinetic power of the outflow.}
\begin{center}
\begin{tabular}{cccccc}\hline
$\rm\alpha$ & $\rm M_{H_2}$ & $\rm M_{H_2}$ & $\rm M_{H_2}$ & $\rm \dot{M}$ & $\rm \dot{E}$ \\
  & (ring) & (circumnuclear) & (outflow) & & \\
(K \kmps pc\pp{2})$^{-1}$ & ($\times$ 10\pp{7} \Msun) & ($\times$ 10\pp{6} \Msun) & ($\times$ 10\pp{6} \Msun) & (\Msun yr\p{1}) & ($\times$ 10\pp{41} erg s\p{1}) \\
 (1) & (2) & (3) & (4) & (5) & (6)\\
\hline
3.4  &  (18.0$\pm$0.7)  & -     &  -     & -     & -\\
0.8  &  -               & (15.7$\pm$1.6)  &  (11.7$\pm$ 1.6) &  (10.5$\pm$1.4) & (4.1$\pm$ 0.6)\\
0.34 &  -               & (6.7 $\pm$ 0.7) &  (5.0$\pm$ 0.7)  & (4.5$\pm$0.6)  & (1.75$\pm$ 0.24)\\
\hline
\end{tabular}    
\end{center}
The columns are: (1) The CO to H$_2$ conversion factor: for Milky Way-like galaxies\cite{Daddi2010}, $\alpha =3.4$ \Msun (K \kmps pc\pp{2})$^{-1}$; for ULIRGs where the ISM is more turbulent due to star formation activity\cite{Downes1998}, $\alpha =0.8$ \Msun (K \kmps pc\pp{2})$^{-1}$; for highly turbulent optically thin gas\cite{Bolatto2013}, $\alpha = 0.34$ \Msun (K \kmps pc\pp{2})$^{-1}$; (2),(3) and (4) The molecular gas masses of the large ring, the circumnuclear disc and the outflow respectively corresponding to the value of $\alpha$ in (1); (5) The mass outflow rate corresponding to (1), (6) The kinetic power of the outflow, also corresponding to the $\alpha$ listed in (1).\\ $S_{CO} \Delta v$ values for the ring, circumnuclear gas and the outflow are: (3.8$\pm$0.2) Jy \kmps, (1.5$\pm$0.2) Jy \kmps, and (1.1$\pm$0.2) Jy \kmps\ respectively.
\label{tab:outflow_parameters}
\end{table}

\clearpage



\begin{thebibliography}{10}
\expandafter\ifx\csname url\endcsname\relax
  \def\url#1{\texttt{#1}}\fi
\expandafter\ifx\csname urlprefix\endcsname\relax\def\urlprefix{URL }\fi
\providecommand{\bibinfo}[2]{#2}
\providecommand{\eprint}[2][]{\url{#2}}

\bibitem{McNamara2012}
\bibinfo{author}{McNamara, B.~R.} \& \bibinfo{author}{Nulsen, P. E.~J.}
\newblock \bibinfo{title}{{Mechanical feedback from active galactic nuclei in
  galaxies, groups and clusters}}.
\newblock \emph{\bibinfo{journal}{\njp}}
  \textbf{\bibinfo{volume}{14}}, \bibinfo{pages}{55023} (\bibinfo{year}{2012}).

\bibitem{Veilleux2020}
\bibinfo{author}{Veilleux, S.}, \bibinfo{author}{Maiolino, R.},
  \bibinfo{author}{Bolatto, A.~D.} \& \bibinfo{author}{Aalto, S.}
\newblock \bibinfo{title}{{Cool outflows in galaxies and their implications}}.
\newblock \emph{\bibinfo{journal}{\aapr}}
  \textbf{\bibinfo{volume}{28}}, \bibinfo{pages}{2} (\bibinfo{year}{2020}).

\bibitem{Sutherland2007}
\bibinfo{author}{Sutherland, R.~S.} \& \bibinfo{author}{Bicknell, G.~V.}
\newblock \bibinfo{title}{{Interactions of a Light Hypersonic Jet with a
  Nonuniform Interstellar Medium}}.
\newblock \emph{\bibinfo{journal}{\apjs}}
  \textbf{\bibinfo{volume}{173}}, \bibinfo{pages}{37--69}
  (\bibinfo{year}{2007}).

\bibitem{Wagner2012}
\bibinfo{author}{Wagner, A.~Y.}, \bibinfo{author}{Bicknell, G.~V.} \&
  \bibinfo{author}{Umemura, M.}
\newblock \bibinfo{title}{{Driving outflows with relativistic jets and the
  dependence of active galactic nucleus feedback efficiency on interstellar
  medium inhomogeneity}}.
\newblock \emph{\bibinfo{journal}{\apj}}
  \textbf{\bibinfo{volume}{757}}, \bibinfo{pages}{136}
  (\bibinfo{year}{2012}).

\bibitem{Mukherjee2016}
\bibinfo{author}{Mukherjee, D.}, \bibinfo{author}{Bicknell, G.~V.},
  \bibinfo{author}{Sutherland, R.} \& \bibinfo{author}{Wagner, A.}
\newblock \bibinfo{title}{{Relativistic jet feedback in high-redshift galaxies
  - I. Dynamics}}.
\newblock \emph{\bibinfo{journal}{\mnras
  }} \textbf{\bibinfo{volume}{461}}, \bibinfo{pages}{967--983}
  (\bibinfo{year}{2016}).

\bibitem{Mukherjee18b}
\bibinfo{author}{Mukherjee, D.}, \bibinfo{author}{Bicknell, G.~V.},
  \bibinfo{author}{Wagner, A.~Y.}, \bibinfo{author}{Sutherland, R.~S.} \&
  \bibinfo{author}{Silk, J.}
\newblock \bibinfo{title}{{Relativistic jet feedback - III. Feedback on gas
  discs}}.
\newblock \emph{\bibinfo{journal}{\mnras
  }} \textbf{\bibinfo{volume}{479}}, \bibinfo{pages}{5544--5566}
  (\bibinfo{year}{2018}).

\bibitem{Brienza2018}
\bibinfo{author}{Brienza, M.} \emph{et~al.}
\newblock \bibinfo{title}{{Astrophysics Duty cycle of the radio galaxy B2 0258
  + 35}}.
\newblock \emph{\bibinfo{journal}{\aap}}
  \textbf{\bibinfo{volume}{45}}, \bibinfo{pages}{1--12} (\bibinfo{year}{2018}).

\bibitem{Giroletti2005}
\bibinfo{author}{Giroletti, M.}, \bibinfo{author}{Giovannini, G.} \&
  \bibinfo{author}{Taylor, G.~B.}
\newblock \bibinfo{title}{{Low power compact radio galaxies at high angular
  resolution}}.
\newblock \emph{\bibinfo{journal}{\aap}}
  \textbf{\bibinfo{volume}{441}}, \bibinfo{pages}{89--101}
  (\bibinfo{year}{2005}).

\bibitem{Ho2009}
\bibinfo{author}{Ho, L.~C.}
\newblock \bibinfo{title}{{Origin and dynamical support of ionized gas in
  galaxy bulges}}.
\newblock \emph{\bibinfo{journal}{\apj}}
  \textbf{\bibinfo{volume}{699}}, \bibinfo{pages}{638--648}
  (\bibinfo{year}{2009}).

\bibitem{Murthy19}
\bibinfo{author}{Murthy, S.} \emph{et~al.}
\newblock \bibinfo{title}{{Feedback from low-luminosity radio galaxies: B2
  0258+35}}.
\newblock \emph{\bibinfo{journal}{\aap}}
  \textbf{\bibinfo{volume}{629}}, \bibinfo{pages}{A58} (\bibinfo{year}{2019}).

\bibitem{Shulevski12}
\bibinfo{author}{Shulevski, A.}, \bibinfo{author}{Morganti, R.},
  \bibinfo{author}{Oosterloo, T.} \& \bibinfo{author}{Struve, C.}
\newblock \bibinfo{title}{{Recurrent radio emission and gas supply: the radio
  galaxy B2 0258+35}}.
\newblock \emph{\bibinfo{journal}{\aap}}
  \textbf{\bibinfo{volume}{545}}, \bibinfo{pages}{A91} (\bibinfo{year}{2012}).

\bibitem{Struve10c}
\bibinfo{author}{Struve, C.}, \bibinfo{author}{Oosterloo, T.},
  \bibinfo{author}{Sancisi, R.}, \bibinfo{author}{Morganti, R.} \&
  \bibinfo{author}{Emonts, B.~H.~C.}
\newblock \bibinfo{title}{{Cold gas in massive early-type galaxies: the case of
  NGC 1167}}.
\newblock \emph{\bibinfo{journal}{\aap}}
  \textbf{\bibinfo{volume}{523}}, \bibinfo{pages}{A75} (\bibinfo{year}{2010}).

\bibitem{Oosterloo2017}
\bibinfo{author}{Oosterloo, T.} \emph{et~al.}
\newblock \bibinfo{title}{{Properties of the molecular gas in the fast outflow
  in the Seyfert galaxy IC 5063}}.
\newblock \emph{\bibinfo{journal}{\aap}}
  \textbf{\bibinfo{volume}{608}}, \bibinfo{pages}{A38} (\bibinfo{year}{2017}).

\bibitem{Oosterloo2019}
\bibinfo{author}{Oosterloo, T.} \emph{et~al.}
\newblock \bibinfo{title}{{ALMA observations of PKS 1549-79: a case of feeding
  and feedback in a young radio quasar}}.
\newblock \emph{\bibinfo{journal}{\aap}}
  \textbf{\bibinfo{volume}{632}}, \bibinfo{pages}{A66} (\bibinfo{year}{2019}).

\bibitem{Holt2006}
\bibinfo{author}{Holt, J.} \emph{et~al.}
\newblock \bibinfo{title}{{The co-evolution of the obscured quasar PKS 1549-79
  and its host galaxy: evidence for a high accretion rate and warm outflow}}.
\newblock \emph{\bibinfo{journal}{\mnras
  }} \textbf{\bibinfo{volume}{370}}, \bibinfo{pages}{1633--1650}
  (\bibinfo{year}{2006}).

\bibitem{Willott99}
\bibinfo{author}{Willott, C.~J.}, \bibinfo{author}{Rawlings, S.},
  \bibinfo{author}{Blundell, K.~M.} \& \bibinfo{author}{Lacy, M.}
\newblock \bibinfo{title}{{The emission line-radio correlation for radio
  sources using the 7C Redshift Survey}}.
\newblock \emph{\bibinfo{journal}{\mnras
  }} \textbf{\bibinfo{volume}{309}}, \bibinfo{pages}{1017--1033}
  (\bibinfo{year}{1999}).

\bibitem{Cavagnolo10}
\bibinfo{author}{Cavagnolo, K.~W.} \emph{et~al.}
\newblock \bibinfo{title}{{A Relationship Between AGN Jet Power and Radio
  Power}}.
\newblock \emph{\bibinfo{journal}{\apj}}
  \textbf{\bibinfo{volume}{720}}, \bibinfo{pages}{1066--1072}
  (\bibinfo{year}{2010}).

\bibitem{Mukherjee18a}
\bibinfo{author}{Mukherjee, D.} \emph{et~al.}
\newblock \bibinfo{title}{{The jet-ISM interactions in IC 5063}}.
\newblock \emph{\bibinfo{journal}{\mnras
  }} \textbf{\bibinfo{volume}{476}}, \bibinfo{pages}{80--95}
  (\bibinfo{year}{2018}).

\bibitem{Capetti1996}
\bibinfo{author}{Capetti, A.}, \bibinfo{author}{Axon, D.~J.},
  \bibinfo{author}{Macchetto, F.}, \bibinfo{author}{Sparks, W.~B.} \&
  \bibinfo{author}{Boksenberg, A.}
\newblock \bibinfo{title}{{Radio Outflows and the Origin of the Narrow-Line
  Region in Seyfert Galaxies}}.
\newblock \emph{\bibinfo{journal}{\apj}}
  \textbf{\bibinfo{volume}{469}}, \bibinfo{pages}{554} (\bibinfo{year}{1996}).

\bibitem{Wilson1999}
\bibinfo{author}{Wilson, A.~S.} \& \bibinfo{author}{Raymond, J.~C.}
\newblock \bibinfo{title}{{Do Jet-driven Shocks Ionize the Narrow-Line Regions
  of Seyfert Galaxies?}}
\newblock \emph{\bibinfo{journal}{\apj}}
  \textbf{\bibinfo{volume}{513}}, \bibinfo{pages}{L115--L118}
  (\bibinfo{year}{1999}).

\bibitem{Morganti15}
\bibinfo{author}{Morganti, R.}, \bibinfo{author}{Oosterloo, T.},
  \bibinfo{author}{Oonk, J.~B.~R.}, \bibinfo{author}{Frieswijk, W.} \&
  \bibinfo{author}{Tadhunter, C.}
\newblock \bibinfo{title}{{The fast molecular outflow in the Seyfert galaxy IC
  5063 as seen by ALMA}}.
\newblock \emph{\bibinfo{journal}{\aapr}}
  \textbf{\bibinfo{volume}{580}}, \bibinfo{pages}{A1} (\bibinfo{year}{2015}).

\bibitem{Garcia-Burillo2014}
\bibinfo{author}{Garc{\'{i}}a-Burillo, S.} \emph{et~al.}
\newblock \bibinfo{title}{{Molecular line emission in NGC 1068 imaged with ALMA
  : I. An AGN-driven outflow in the dense molecular gas}}.
\newblock \emph{\bibinfo{journal}{\aap}}
  \textbf{\bibinfo{volume}{567}}, \bibinfo{pages}{1--24}
  (\bibinfo{year}{2014}).

\bibitem{Venturi2021}
\bibinfo{author}{Venturi, G.} \emph{et~al.}
\newblock \bibinfo{title}{{MAGNUM survey: Compact jets causing large turmoil in
  galaxies. Enhanced line widths perpendicular to radio jets as tracers of
  jet-ISM interaction}}.
\newblock \emph{\bibinfo{journal}{\aap}}
  \textbf{\bibinfo{volume}{648}}, \bibinfo{pages}{A17} (\bibinfo{year}{2021}).

\bibitem{Jarvis2019}
\bibinfo{author}{Jarvis, M.~E.} \emph{et~al.}
\newblock \bibinfo{title}{{Prevalence of radio jets associated with galactic
  outflows and feedback from quasars}}.
\newblock \emph{\bibinfo{journal}{\mnras
  }} \textbf{\bibinfo{volume}{485}}, \bibinfo{pages}{2710--2730}
  (\bibinfo{year}{2019}).

\bibitem{Husemann2019}
\bibinfo{author}{Husemann, B.} \emph{et~al.}
\newblock \bibinfo{title}{{The Close AGN Reference Survey (CARS): A massive
  multi-phase outflow impacting the edge-on galaxy HE 1353-1917}}.
\newblock \emph{\bibinfo{journal}{\aap}}
  \textbf{\bibinfo{volume}{627}}, \bibinfo{pages}{A53} 
  (\bibinfo{year}{2019}).

\bibitem{Alatalo11}
\bibinfo{author}{Alatalo, K.} \emph{et~al.}
\newblock \bibinfo{title}{{Discovery of an Active Galactic Nucleus Driven
  Molecular Outflow in the Local Early-type Galaxy NGC 1266}}.
\newblock \emph{\bibinfo{journal}{\apj}}
  \textbf{\bibinfo{volume}{735}}, \bibinfo{pages}{88} (\bibinfo{year}{2011}).

\bibitem{Combes2013}
\bibinfo{author}{Combes, F.} \emph{et~al.}
\newblock \bibinfo{title}{{ALMA observations of feeding and feedback in nearby
  Seyfert galaxies: an AGN-driven outflow in NGC 1433}}.
\newblock \emph{\bibinfo{journal}{\aap}}
  \textbf{\bibinfo{volume}{558}}, \bibinfo{pages}{A124} (\bibinfo{year}{2013}).

\bibitem{Riffel2014}
\bibinfo{author}{Riffel, R.~A.}, \bibinfo{author}{Storchi-Bergmann, T.} \&
  \bibinfo{author}{Riffel, R.}
\newblock \bibinfo{title}{{An Outflow Perpendicular to the Radio Jet in the
  Seyfert Nucleus of NGC 5929}}.
\newblock \emph{\bibinfo{journal}{\apj}}
  \textbf{\bibinfo{volume}{780}}, \bibinfo{pages}{L24} (\bibinfo{year}{2014}).

\bibitem{Rodriguez-Ardila2017}
\bibinfo{author}{Rodr{\'{i}}guez-Ardila, A.} \emph{et~al.}
\newblock \bibinfo{title}{{Powerful outflows in the central parsecs of the
  low-luminosity active galactic nucleus NGC 1386}}.
\newblock \emph{\bibinfo{journal}{\mnras
  }} \textbf{\bibinfo{volume}{470}}, \bibinfo{pages}{2845--2860}
  (\bibinfo{year}{2017}).

\bibitem{Fabbiano18b}
\bibinfo{author}{Fabbiano, G.} \emph{et~al.}
\newblock \bibinfo{title}{{Deep Chandra Observations of ESO 428-G01. III.
  High-resolution Spectral Imaging of the Ionization Cone and Radio Jet
  Region}}.
\newblock \emph{\bibinfo{journal}{\apj}}
  \textbf{\bibinfo{volume}{865}}, \bibinfo{pages}{83} (\bibinfo{year}{2018}).

\bibitem{Gomes2016WarmGalaxies}
\bibinfo{author}{Gomes, J.~M.} \emph{et~al.}
\newblock \bibinfo{title}{{Warm ionized gas in CALIFA early-type galaxies. 2D
  emission-line patterns and kinematics for 32 galaxies}}.
\newblock \emph{\bibinfo{journal}{\aap}}
  \textbf{\bibinfo{volume}{588}}, \bibinfo{pages}{A68} (\bibinfo{year}{2016}).

\bibitem{Mukherjee2017}
\bibinfo{author}{Mukherjee, D.}, \bibinfo{author}{Bicknell, G.~V.},
  \bibinfo{author}{Sutherland, R.} \& \bibinfo{author}{Wagner, A.}
\newblock \bibinfo{title}{{Erratum: Relativistic jet feedback in high-redshift
  galaxies I. Dynamics}}.
\newblock \emph{\bibinfo{journal}{\mnras
  }} \textbf{\bibinfo{volume}{471}}, \bibinfo{pages}{2790--2800}
  (\bibinfo{year}{2017}).

\bibitem{Wagner2011RelativisticGalaxies}
\bibinfo{author}{Wagner, A.~Y.} \& \bibinfo{author}{Bicknell, G.~V.}
\newblock \bibinfo{title}{{Relativistic Jet Feedback in Evolving Galaxies}}.
\newblock \emph{\bibinfo{journal}{\apj}}
  \textbf{\bibinfo{volume}{728}}, \bibinfo{pages}{29} (\bibinfo{year}{2011}).

\bibitem{Gaibler2011AsymmetriesInteraction}
\bibinfo{author}{Gaibler, V.}, \bibinfo{author}{Khochfar, S.} \&
  \bibinfo{author}{Krause, M.}
\newblock \bibinfo{title}{{Asymmetries in extragalactic double radio sources:
  clues from 3D simulations of jet-disc interaction}}.
\newblock \emph{\bibinfo{journal}{\mnras
  }} \textbf{\bibinfo{volume}{411}}, \bibinfo{pages}{155--161}
  (\bibinfo{year}{2011}).

\bibitem{Best2005}
\bibinfo{author}{Best, P.~N.}, \bibinfo{author}{Kauffmann, G.},
  \bibinfo{author}{Heckman, T.~M.} \& \bibinfo{author}{Ivezi{\'{c}}, Z.}
\newblock \bibinfo{title}{{A sample of radio-loud active galactic nuclei in the
  Sloan Digital Sky Survey}}.
\newblock \emph{\bibinfo{journal}{\mnras
  }} \textbf{\bibinfo{volume}{362}}, \bibinfo{pages}{9--24}
  (\bibinfo{year}{2005}).

\bibitem{Sabater2019}
\bibinfo{author}{Sabater, J.} \emph{et~al.}
\newblock \bibinfo{title}{{The LoTSS view of radio AGN in the local Universe:
  The most massive galaxies are always switched on}}.
\newblock \emph{\bibinfo{journal}{\aap}}
  \textbf{\bibinfo{volume}{622}}, \bibinfo{pages}{1--14}
  (\bibinfo{year}{2019}).

\bibitem{Morganti2018TheAbsorption}
\bibinfo{author}{Morganti, R.} \& \bibinfo{author}{Oosterloo, T.}
\newblock \bibinfo{title}{{The interstellar and circumnuclear medium of active
  nuclei traced by H i 21 cm absorption}}.
\newblock \emph{\bibinfo{journal}{\aap Review}}
  \textbf{\bibinfo{volume}{26}}, \bibinfo{pages}{4} (\bibinfo{year}{2018}).

\bibitem{Genel2014IntroducingTime}
\bibinfo{author}{Genel, S.} \emph{et~al.}
\newblock \bibinfo{title}{{Introducing the Illustris project: the evolution of
  galaxy populations across cosmic time}}.
\newblock \emph{\bibinfo{journal}{\mnras
  }} \textbf{\bibinfo{volume}{445}}, \bibinfo{pages}{175--200}
  (\bibinfo{year}{2014}).

\bibitem{Sijacki2015TheTime}
\bibinfo{author}{Sijacki, D.} \emph{et~al.}
\newblock \bibinfo{title}{{The Illustris simulation: the evolving population of
  black holes across cosmic time}}.
\newblock \emph{\bibinfo{journal}{\mnras
  }} \textbf{\bibinfo{volume}{452}}, \bibinfo{pages}{575--596}
  (\bibinfo{year}{2015}).

\bibitem{Schaye2015TheEnvironments}
\bibinfo{author}{Schaye, J.} \emph{et~al.}
\newblock \bibinfo{title}{{The EAGLE project: simulating the evolution and
  assembly of galaxies and their environments}}.
\newblock \emph{\bibinfo{journal}{\mnras
  }} \textbf{\bibinfo{volume}{446}}, \bibinfo{pages}{521--554}
  (\bibinfo{year}{2015}).

\bibitem{Crain2015TheVariations}
\bibinfo{author}{Crain, R.~A.} \emph{et~al.}
\newblock \bibinfo{title}{{The EAGLE simulations of galaxy formation:
  calibration of subgrid physics and model variations}}.
\newblock \emph{\bibinfo{journal}{\mnras
  }} \textbf{\bibinfo{volume}{450}}, \bibinfo{pages}{1937--1961}
  (\bibinfo{year}{2015}).

\bibitem{Weinberger17}
\bibinfo{author}{Weinberger, R.} \emph{et~al.}
\newblock \bibinfo{title}{{Simulating galaxy formation with black hole driven
  thermal and kinetic feedback}}.
\newblock \emph{\bibinfo{journal}{\mnras
  }} \textbf{\bibinfo{volume}{465}}, \bibinfo{pages}{3291--3308}
  (\bibinfo{year}{2017}).

\bibitem{Dubois12}
\bibinfo{author}{Dubois, Y.}, \bibinfo{author}{Devriendt, J.},
  \bibinfo{author}{Slyz, A.} \& \bibinfo{author}{Teyssier, R.}
\newblock \bibinfo{title}{{Self-regulated growth of supermassive black holes by
  a dual jet-heating active galactic nucleus feedback mechanism: methods, tests
  and implications for cosmological simulations}}.
\newblock \emph{\bibinfo{journal}{\mnras
  }} \textbf{\bibinfo{volume}{420}}, \bibinfo{pages}{2662--2683}
  (\bibinfo{year}{2012}).

\bibitem{Dubois2013}
\bibinfo{author}{Dubois, Y.}, \bibinfo{author}{Gavazzi, R.},
  \bibinfo{author}{Peirani, S.} \& \bibinfo{author}{Silk, J.}
\newblock \bibinfo{title}{{AGN-driven quenching of star formation:
  Morphological and dynamical implications for early-type galaxies}}.
\newblock \emph{\bibinfo{journal}{\mnras
  }} \textbf{\bibinfo{volume}{433}}, \bibinfo{pages}{3297--3313}
  (\bibinfo{year}{2013}).

\bibitem{Talbot2021}
\bibinfo{author}{Talbot, R.~Y.}, \bibinfo{author}{Bourne, M.~A.} \&
  \bibinfo{author}{Sijacki, D.}
\newblock \bibinfo{title}{{Blandford-Znajek jets in galaxy formation
  simulations: Method and implementation}}.
\newblock \emph{\bibinfo{journal}{\mnras
  }} \textbf{\bibinfo{volume}{504}}, \bibinfo{pages}{3619-3650} 
  (\bibinfo{year}{2021}).

\bibitem{Talbot2021Blandford-ZnajekJets}
\bibinfo{author}{Talbot, R.~Y.}, \bibinfo{author}{Sijacki, D.} \&
  \bibinfo{author}{Bourne, M.~A.}
\newblock \bibinfo{title}{{Blandford-Znajek jets in galaxy formation
  simulations: exploring the diversity of outflows produced by spin-driven AGN
  jets}}.
  \newblock \emph{\bibinfo{journal}{arXiv e-prints
  }} \bibinfo{pages}{arXiv:2111.01801} 
  (\bibinfo{year}{2021}).

\bibitem{OSullivan2015}
\bibinfo{author}{O'Sullivan, E.} \emph{et~al.}
\newblock \bibinfo{title}{{Cold gas in group-dominant elliptical galaxies}}.
\newblock \emph{\bibinfo{journal}{\aap}}
  \textbf{\bibinfo{volume}{573}}, \bibinfo{pages}{A111} (\bibinfo{year}{2015}).

\bibitem{Bolatto2017}
\bibinfo{author}{Bolatto, A.~D.} \emph{et~al.}
\newblock \bibinfo{title}{{The EDGE-CALIFA Survey: Interferometric Observations
  of 126 Galaxies with CARMA}}.
\newblock \emph{\bibinfo{journal}{\apj}}
  \textbf{\bibinfo{volume}{846}}, \bibinfo{pages}{159} (\bibinfo{year}{2017}).

\bibitem{Gomes2016Spiral-likeGalaxies}
\bibinfo{author}{Gomes, J.~M.} \emph{et~al.}
\newblock \bibinfo{title}{{Spiral-like star-forming patterns in CALIFA
  early-type galaxies}}.
\newblock \emph{\bibinfo{journal}{\aap}}
  \textbf{\bibinfo{volume}{585}}, \bibinfo{pages}{A92} (\bibinfo{year}{2016}).

\bibitem{Collaboration2014PlanckResults}
\bibinfo{author}{Collaboration, P.} \emph{et~al.}
\newblock \bibinfo{title}{{Planck 2013 results. I. Overview of products and
  scientific results}}.
\newblock \emph{\bibinfo{journal}{\aap}}
  \textbf{\bibinfo{volume}{571}}, \bibinfo{pages}{A1} (\bibinfo{year}{2014}).

\bibitem{Greisen2003AIPSVLBA}
\bibinfo{author}{Greisen, E.~W.}
\newblock \bibinfo{title}{{AIPS, the VLA, and the VLBA}}.
\newblock In \emph{\bibinfo{booktitle}{Information Handling in Astronomy -
  Historical Vistas}}, vol. \bibinfo{volume}{285}, \bibinfo{pages}{109}
  (\bibinfo{address}{AA(National Radio Astronomy Observatory)},
  \bibinfo{year}{2003}).

\bibitem{Solomon2005}
\bibinfo{author}{Solomon, P.~M.} \& \bibinfo{author}{Vanden~Bout, P.~A.}
\newblock \bibinfo{title}{{Molecular Gas at High Redshift}}.
\newblock \emph{\bibinfo{journal}{\araa}}
  \textbf{\bibinfo{volume}{43}}, \bibinfo{pages}{677--725}
  (\bibinfo{year}{2005}).

\bibitem{Daddi2010}
\bibinfo{author}{Daddi, E.} \emph{et~al.}
\newblock \bibinfo{title}{{Different Star Formation Laws for Disks Versus
  Starbursts at Low and High Redshifts}}.
\newblock \emph{\bibinfo{journal}{\apj}}
  \textbf{\bibinfo{volume}{714}}, \bibinfo{pages}{L118--L122}
  (\bibinfo{year}{2010}).

\bibitem{Downes1998}
\bibinfo{author}{Downes, D.} \& \bibinfo{author}{Solomon, P.~M.}
\newblock \bibinfo{title}{{Rotating Nuclear Rings and Extreme Starbursts in
  Ultraluminous Galaxies}}.
\newblock \emph{\bibinfo{journal}{\apj}}
  \textbf{\bibinfo{volume}{507}}, \bibinfo{pages}{615--654}
  (\bibinfo{year}{1998}).

\bibitem{Bolatto2013}
\bibinfo{author}{Bolatto, A.~D.}, \bibinfo{author}{Wolfire, M.} \&
  \bibinfo{author}{Leroy, A.~K.}
\newblock \bibinfo{title}{{The CO-to-H<SUB>2</SUB> Conversion Factor}}.
\newblock \emph{\bibinfo{journal}{\araa}}
  \textbf{\bibinfo{volume}{51}}, \bibinfo{pages}{207--268}
  (\bibinfo{year}{2013}).

\bibitem{Harrison2018}
\bibinfo{author}{Harrison, C.~M.} \emph{et~al.}
\newblock \bibinfo{title}{{AGN outflows and feedback twenty years on}}.
\newblock \emph{\bibinfo{journal}{\nastro}}
  \textbf{\bibinfo{volume}{2}}, \bibinfo{pages}{198--205}
  (\bibinfo{year}{2018}).

\bibitem{Rupke2013TheMergers}
\bibinfo{author}{Rupke, D. S.~N.} \& \bibinfo{author}{Veilleux, S.}
\newblock \bibinfo{title}{{The Multiphase Structure and Power Sources of
  Galactic Winds in Major Mergers}}.
\newblock \emph{\bibinfo{journal}{\apj}}
  \textbf{\bibinfo{volume}{768}}, \bibinfo{pages}{75} (\bibinfo{year}{2013}).

\bibitem{Ho1997}
\bibinfo{author}{Ho, L.~C.}, \bibinfo{author}{Filippenko, A.~V.},
  \bibinfo{author}{Sargent, W. L.~W.} \& \bibinfo{author}{Peng, C.~Y.}
\newblock \bibinfo{title}{{A Search for ``Dwarf'' Seyfert Nuclei. IV. Nuclei
  with Broad H{$\alpha$} Emission}}.
\newblock \emph{\bibinfo{journal}{\apjs}}
  \textbf{\bibinfo{volume}{112}}, \bibinfo{pages}{391--414}
  (\bibinfo{year}{1997}).

\bibitem{Akylas2009XMM-NewtonDistribution}
\bibinfo{author}{Akylas, A.} \& \bibinfo{author}{Georgantopoulos, I.}
\newblock \bibinfo{title}{{XMM-Newton observations of Seyfert galaxies from the
  Palomar spectroscopic survey: the X-ray absorption distribution}}.
\newblock \emph{\bibinfo{journal}{\aap}}
  \textbf{\bibinfo{volume}{500}}, \bibinfo{pages}{999--1012}
  (\bibinfo{year}{2009}).

\bibitem{Panessa2006OnGalaxies}
\bibinfo{author}{Panessa, F.} \emph{et~al.}
\newblock \bibinfo{title}{{On the X-ray, optical emission line and black hole
  mass properties of local Seyfert galaxies}}.
\newblock \emph{\bibinfo{journal}{\aap}}
  \textbf{\bibinfo{volume}{455}}, \bibinfo{pages}{173--185}
  (\bibinfo{year}{2006}).

\bibitem{Godfrey2016}
\bibinfo{author}{Godfrey, L. E.~H.} \& \bibinfo{author}{Shabala, S.~S.}
\newblock \bibinfo{title}{{Mutual distance dependence drives the observed
  jet-power-radio-luminosity scaling relations in radio galaxies}}.
\newblock \emph{\bibinfo{journal}{\mnras
  }} \textbf{\bibinfo{volume}{456}}, \bibinfo{pages}{1172--1184}
  (\bibinfo{year}{2016}).

\end{thebibliography}

\end{document}